\documentstyle[12pt,psfig,draft]{nature}

\def\ra#1#2#3{#1$^{\rm h}$#2$^{\rm m}$#3$^{\rm s}$}
\def\dec#1#2#3{$#1^\circ#2'#3''$}   

\begin{document}

\title{\large \bf Discovery of Radio Emission from the Brown Dwarf LP944$-$20}

\author{
E. Berger\affiliation[1]{Division of Physics, Mathematics \&\ Astronomy 105-24,
         California Institute of Technology, Pasadena, CA 91125,USA},
S. Ball\affiliation[2]{Department of Physics, New Mexico Tech, PO Box 3394 c/s, Socorro, NM 87801},
K. M. Becker\affiliation[3]{Department of Physics, Oberlin College, Oberlin, OH 44074},
M. Clarke\affiliation[4]{Department of Physics, Carleton College, 300 N. College St., Northfield, MN 55057},
D. A. Frail\affiliation[5]{National Radio Astronomy Observatory, P. O. Box O, Socorro, NM 87801, USA},
T. A. Fukuda\affiliation[6]{Department of Physics \& Astronomy, University of Denver, 2112
         E. Wesley Ave., Denver, CO 80208},
I. M. Hoffman\affiliation[7]{Department of Physics \& Astronomy, University of New Mexico, 800 Yale Blvd. 
         NE, Albuquerque, NM 87131},
S. R. Kulkarni$^1$,
R. Mellon\affiliation[8]{Department of Astronomy, 525 Davey Lab., The Pennsylvania
         State University, University Park, PA 16082},
E. Momjian\affiliation[9]{Department of Physics \& Astronomy, 177 chem-physics building,
         University of Kentucky, Lexington, KY 40506},
N. W. Murphy\affiliation[10]{Department of Physics, Amherst College, Amherst, MA 01002-5000},
S. H. Teng\affiliation[11]{Department of Astronomy, Univeristy of Maryland, College Park, MD 20742},
T. Woodruff\affiliation[12]{Department of Physics, Southwestern University, S.U. Box 7263, Georgetown,
         TX 78626},
B. A. Zauderer\affiliation[13]{Department of Physics \& Astronomy, Agnes Scott College, Decatur, GA 30030},
R. T. Zavala\affiliation[14]{Department of Astronomy, New Mexico State University, MSC 4500 PO Box 30001, 
         Las Cruces, NM 88003}
}
\dates{\today}{} 
\headertitle{}
\mainauthor{}
\smallskip

\summary{
Brown dwarfs are classified as objects which are not massive enough to sustain nuclear fusion of 
hydrogen, and are distinguished from planets by their ability to burn deuterium\cite{nbc+99}.  
Old ($>\!\!10$ Myr) brown dwarfs are expected to possess short-lived magnetic fields\cite{dyr93} and, since they no 
longer generate energy from collapse and accretion, weak radio and X-ray emitting coronae.  Several efforts have 
been undertaken in the past to 
detect chromospheric activity from the brown dwarf LP944$-$20 at X-ray~\cite{nbc+99}$^,$~\cite{rbm+00} and 
optical~\cite{t98}$^,$~\cite{tr98}$^,$~\cite{tt99}$^,$~\cite{khi97} 
wavelengths, but only recently an X-ray flare from this object was detected\cite{rbm+00}.  
Here we report on the discovery of quiescent and flaring 
radio emission from this source, which represents the first detection of persistent radio emission from a brown 
dwarf, with luminosities that are several orders of magnitude larger than predicted from an empirical 
relation~\cite{gb93}$^,$~\cite{bg94} between the X-ray and radio luminosities of many stellar types.
We show in the context of synchrotron emission, that LP944$-$20 possesses an unusually weak magnetic field in 
comparison to active dwarf M stars,\cite{sl85}$^,$\cite{hsr91} which might explain the null results from previous
optical and X-ray observations of this source, and the deviation from the empirical relations.}

\maketitle

\noindent{\it This paper has been submitted to Nature.
  You are free to use the results here for the purpose of your
  research.  In accordance with the editorial policy of Nature, we
  request that you not discuss this result in the popular press. If
  you have any question or need clarifications please contact Edo
  Berger, ejb@astro.caltech.edu.}

\vspace{0.2in}

Following the detection of an X-ray flare from the brown dwarf LP944$-$20 on 1999 December 15 
by the {\it Chandra} X-ray Observatory \cite{rbm+00}, we observed the X-ray position of this source 
with the Very Large Array (VLA) on 2000 July 27.54 UT, and detected a radio source.  We tested the radio 
emission for variability by imaging subsets of the observation and constructing
a light-curve, and found a bright flare, as well as quiescent emission (see Figure~1).  In subsequent 
observations we detected two additional flares, and quiescent emission at 4.9 and 8.5 GHz.  A log of all 
observations, the details of the data reduction, and results can be found in Table~\ref{tab:vlaobs} and 
Figure~1.  The quiescent emission at the location of the
source was detected at the $2-3 \sigma$ level in individual measurements, but with a level of 2.5 and 5.5 $\sigma$
for the average of all measurements at 4.9 and 8.5 GHz, respectively.  Since we know the precise location of the 
source, these levels represent a significant detection of quiescent emission.  

This set of observations comprises the first detection of persistent quiescent and flaring 
emission from a bona-fide brown dwarf.\cite{nbc+99}$^,$\cite{kll99}  Moreover, the levels of emission --- 
approximately 80 $\mu$Jy in quiescence and 2 mJy at peak-flare --- are highly 
unusual in the context of stellar coronal emission.  We now seek to interpret the quiescent and flaring emission 
in this framework.  

Coronal activity in a broad range of stars has been studied by Guedel and 
Benz who found\cite{gb93} a simple empirical relation between the quiescent X-ray and radio luminosities, 
$L_R\approx L_X/10^{15.5}$ Hz$^{-1}$.  Using this relation with the 
{\it Chandra}-determined\cite{rbm+00} upper limit on the quiescent X-ray luminosity of LP944$-$20, $L_X<10^{24}$ 
erg sec$^{-1}$ ($0.1-10$ keV band), we find that the expected quiescent radio luminosity is  
$L_R<3.5\times 10^{8}$ erg sec$^{-1}$ Hz$^{-1}$, which translates to a
received flux density of $<0.01$ $\mu$Jy on Earth.  
Therefore, the flux density we measured in quiescence is at least four orders of magnitude larger than predicted.

Benz and Guedel also found\cite{bg94} a similar, but slightly non-linear, relation between X-ray and radio flares 
from a broad range of stars.  The non-linearity means that for flaring emission the ratio $L_X/L_R$ varies with 
the luminosity of the flare.  For the X-ray luminosity detected by {\it Chandra}, 
$L_X \approx 1.2\times 10^{26}$ erg sec$^{-1}$, the relation is $L_X/L_R \approx 10^{16.5}$ Hz, and 
we therefore predict an observed radio flux density, $F_{\nu R} \approx 0.1$ $\mu$Jy. The actual observed value is 
approximately $2\times 10^4$ times higher.  

Even though the X-ray and radio observations 
were not carried simultaneously, X-ray flares bright enough to resolve the discrepancy are not likely.  In order to
conform to the Benz--Guedel relation, the luminosity of the X-ray flares would have to be approximately equal 
to the bolometric luminosity of LP944$-$20.  Such superflaring activity\cite{skd00}$^,$\cite{rs00} has been 
observed in nine {\it solar}-type stars, but with an average recurrence time of years to centuries.  In contrast, 
the average recurrence time for the radio flares is of order several hours (see Table~\ref{tab:lc}).  Moreover, 
with such a short recurrence time, {\it Chandra} would have been expected to detect at least one of these 
superflares during the 13-hour observation of LP944$-$20.  We can therefore conclude that the Benz--Guedel 
relation for flaring emission is severely violated, and the radio emission from LP944$-$20 is highly abnormal
when compared to stellar emission. 

The brightness temperature of the radio emission, approximately $2\times 10^8$ K in quiescence and $4\times 10^9$
K during flares, indicates that the emission mechanism is non-thermal.  Most likely the emission is via 
synchrotron, which comes from relativistic electrons gyrating in a magnetic field and emitting radio waves at 
high harmonics of the electron gyrofrequency.  Another possible mechanism is coherent emission, but the observed 
properties of the quiescent and flaring emission: flare timescale of order minutes, polarization of order 30\%, 
broad-band emission, and $T_b<10^{10}$ K, all point to synchrotron emission;\cite{hsr91}$^,$\cite{d87} for 
coherent emission we would expect a much shorter timescale, a much higher polarization and 
brightness temperature, and narrow-band emission.  For a comprehensive treatment of synchrotron theory 
we refer the reader to the work of Rybicki and Lightman\cite{rl79}.  

To find the spectral properties of the flaring emission, and thus the physical properties of the emission region
around LP944$-$20, we use the observed fraction of circular polarization in the flaring emission at 8.5 GHz, 
$f_{\rm circ}=0.3\pm 0.1$, which matches theoretical predictions\cite{bm90}$^,$\cite{bbg98}$^,$\cite{dm82} for 
optically-thin synchrotron emission at moderate values of the gyrating electrons' minimum Lorentz factor 
($\gamma_{\rm min}$). 
In addition, we find from the simultaneous observation on 2000 August 30.40 UT that the spectral index between 4.9
and 8.5 GHz was $2.1\pm 0.3$, indicating that the emission at 4.9 GHz was optically-thick (see Figure~1c).  We 
therefore conclude that the radio spectrum of the flaring emission peaks at about 8.5 GHz, so that most of the 
energy in the flares at radio wavelengths is released at this frequency.  Thus, we can use $\nu_{\rm peak}=8.5$ 
GHz and $F_\nu (8.5{\rm GHz})\equiv F_{\nu,\rm peak}$ as an excellent approximation for the flare energetics 
(see Table~\ref{tab:lc}). 

The amount of energy released in the flares is estimated by using an exponentially rising and decaying model, and 
solving for the peak flux density and rise and decay times (see Table~\ref{tab:lc}).  We find that the energy 
released in each of the flares at radio wavelengths is approximately the same, ranging from $2-8\times 10^{26}$ 
ergs.  Interestingly, this energy release is several times larger than for the brightest solar 
flares\cite{bbg98}.  From the flare model we also find that the best-fit quiescent flux density is 
$100\pm 35$ $\mu$Jy, and that the probability of no quiescent component is $2\times 10^{-4}$.  This result serves 
to bolster our claim that we detected quiescent emission at a level of 100 $\mu$Jy from LP944$-$20. 

From the fraction of circular polarization we can also estimate the value of $\gamma_{\rm min}$ using the relation 
for synchrotron emission $f_{\rm circ}\approx 3/\gamma_{\rm min}$, which gives $\gamma_{\rm min} \approx 10$.  
Using the more accurate formalism of Dulk and Marsh\cite{dm82}, we get a similar result. 

With the values of $\gamma_{\rm min}$, $\nu_{\rm peak}$, and $F_{\nu,\rm peak}$ we now calculate the 
magnetic field strength and the electron density during a flare from the basic equations of 
synchrotron theory\cite{rl79}, and from the minimum energy argument, which relies on the premise that the energy 
of a source is minimized in equipartition.\cite{bb57}$^,$\cite{r94}  From synchrotron theory we find that the 
magnetic field strength is $B\approx 5$ G, and the total number of electrons is $N_e\approx 2\times 10^{35}$ 
(see Figure~1).  From equipartition we find that the ``equipartition radius''\cite{sr77} of LP944$-$20 
is $\theta_{eq}\approx 75$ $\mu$as, while its physical angular size is $\theta_s\approx 60$ $\mu$as, indicating 
that the energy is released from a thin shell around the source, with a volume, $V\approx 4\times 10^{29}$ cm$^3$.
The electrons' Lorentz factor calculated from equipartition is $\gamma_{\rm min}\approx 10$, which is exactly the 
value we calculated independently from circular polarization, and hence we again find that $B\approx 5$ G and 
$N_e\approx 2\times 10^{35}$.  Using the volume of the emitting shell we calculate an electron number density,
$n_e\approx 6\times 10^5$ cm$^{-3}$.  For comparison, the surface magnetic field of Jupiter\cite{sdj75} is 
approximately 10 G, while in Solar flaring emission\cite{bm90} the magnetic field is $B\sim 80$ G, but 
the electron density and energy are much lower than in the flaring emission from LP944$-$20.  

Similarly, the magnetic fields of some active dwarf M stars\cite{sl85}$^,$\cite{hsr91} reach strengths of 
a few kG, and their X-ray luminosities {\it in quiescence}\cite{flg88}$^,$\cite{grk96}$^,$\cite{fgs93} are of 
order $10^{27}-10^{29}$ erg sec$^{-1}$.  These are markedly different conditions than those in and around
LP944$-$20, as evidenced by the weak magnetic field and non-detectable quiescent X-ray emission, but they may 
help to elucidate several lines of evidence that point to no magnetic field activity in this object.  Extremely 
low levels of H$\alpha$ emission\cite{tr98}, which indicate no chromospheric activity, rapid rotation\cite{tr98},
which indicates no magnetic braking, and relatively old age\cite{t98}, which implies a quenched dynamo 
mechanism, all agree with a magnetic field strength of a few Gauss rather than the much stronger fields observed 
in M dwarfs, the Sun, and even Jupiter.

We finally note that radio synchrotron and X-ray emission are expected in the context of a basic model of coronal 
emission --- the magnetic reconnection process,\cite{bbg98}$^,$\cite{s99} 
and that the previously noted constancy of radio-wavelength energy release during flares appears to indicate 
that magnetic reconnection in LP944$-$20 takes place once the magnetic fields reach a
critical strength, and therefore release approximately the same amount of energy every time.  

It is still not clear why the emission from LP944$-$20 violates the Guedel--Benz relations so severely, but it is 
possible that this violation is tied to the difference in physical conditions in LP944$-$20 relative to M dwarfs, 
namely a very weak magnetic field.  It is therefore crucial that radio and X-ray observations of 
this and other brown dwarfs be undertaken in order to establish whether similar conditions and unusually strong 
radio emission are common to some of these objects.  In particular, 
it is clear that the radio 
regime is highly effective for detailed studies of both the flaring and quiescent 
emission due to the ease and availability of radio observations and the excellent sensitivity of radio 
telescopes.

\clearpage

\begin{acknowledge}
  We thank B. Clark for the allocation of ad hoc VLA time.  We also thank D. E. Gary and R. Sari
  for helpful discussions.  The initial observation of LP944$-$20 was
  undertaken as part of the National Radio Astronomy Observatory (NRAO) VLA Summer 
  Program funded by the National Science Foundation (NSF).
  The National Radio Astronomy Observatory is a facility of the NSF operated under cooperative agreement 
   by Associated Universities, Inc. 
\end{acknowledge}

\clearpage
\begin{table}
\begin{center}
\begin{tabular}{c c c c c}
\hline
\hline
\small Date & Frequency & On-source observation time & Synthesized beam size & $F\pm\sigma$ \\
(UT) & (GHz) & (ksec) & (arcsec) & ($\mu$Jy) \\ \hline
2000 July 27.54 & 8.46 & 3.6 & $26.6 \times 6.4$ & $300\pm 45$  \\
2000 Aug 11.47 & 8.46 & 4.2 & $27.5 \times 7.0$ & $70 \pm 27$  \\
2000 Aug 15.45 & 4.86 & 1.7 & $52.6 \times 12.6$ & $127 \pm 53$  \\
2000 Aug 15.45 & 8.46 & 1.6 & $29.8 \times 7.1$  & $130 \pm 44$ \\
2000 Aug 23.43 & 8.46 & 5.6 & $27.2 \times 7.3$ & $250 \pm 26$  \\
2000 Aug 30.40 & 4.86 & 7.2 & $41.4 \times 12.6$ & $106 \pm 50$  \\
2000 Aug 30.40 & 8.46 & 7.2 & $26.5 \times 7.6$ & $323 \pm 49$  \\ 
2000 Sep 6.39  & 8.46 & 8.8 & $22.9 \times 6.2$ & $70\pm 39$ \\
2000 Sep 6.39  & 22.5 & 8.8 & $10.4 \times 4.4$ & $190\pm 170$ \\ 
2000 Sep 20.34 & 4.86 & 10.5 & $33.9\times 13.7$ & $50\pm 36$ \\
2000 Sep 20.34 & 8.46 & 10.5 & $45.5\times 6.6$ & $140\pm 45$ \\
2000 Sep 27.35 & 8.46 & 10.2 & $22.9 \times 7.4$ & $53\pm 20$ \\ \hline
\end{tabular}
\end{center}
\renewcommand{\baselinestretch}{1.5}
\caption[]{A summary of VLA observations of LP944$-$20, exhibiting both quiescent and flaring emission.
  The columns are (left to right), (1) UT date of the
  start of each observation, (2) observing frequency, (3) On-source observing time,
  (4) Synthesized beam size, and (5) peak flux density at the best fit position of LP944$-$20, 
  with the error given as the root mean square noise on the image.  Single-frequency 
  observations consisted of several on-source scans using the VLA standard continuum 
  mode, with the full 100 MHz bandwidth obtained in two adjacent 50 MHz bands.  The flux 
  density scale was determined using the extra-galactic source 3C48 (J0137+331), and the 
  phase was monitored using the sources J0334$-$401 and J0403$-$360.  Dual-frequency observations were done with
  a split array configuration in which only several antennas
  were used for each observing frequency.  The data were reduced with the 
  Astronomical Image Processing System (AIPS, releases 15APR1999 and 31DEC1999).  \newline
  In the initial observation we detected an object with a flux $F_{8.5{\rm GHz}}=300\pm 45$ $\mu$Jy at 
  $\alpha$(J2000)=\ra{03}{39}{35.3}, $\delta$(J2000)=\dec{-35}{25}{44.0}, with an error of $\sim 1^{''}$ 
  in both $\alpha$ and $\delta$.  We identify our detection with the brown dwarf LP944$-$20 based on the positional
  coincidence between this detection and the X-ray position\cite{rbm+00} provided by {\it Chandra}.}
\label{tab:vlaobs}
\end{table}

\clearpage
\begin{table}
\begin{center}
\begin{tabular}{c c c c c}
\hline
\hline
Parameter & 2000 July 27.5727 & 2000 Aug 23.4407 & 2000 Aug 30.4849 \\ \hline
$F_0$ (mJy)& $1.5 \pm 0.1$ & $2.6\pm 0.2$ & $2.0\pm 0.2$ \\
$\tau_1$ (min) & $3.0^{+4.2}_{-1.5}$ & $6.2^{+1.5}_{-0.8}$ & $20\pm 5$ \\
$\tau_2$ (min) & $5.7^{+2.4}_{-1.6}$ & $1.7^{+0.4}_{-0.3}$ & $5.1^{+2.2}_{-1.5}$ \\
$\delta t_0$ (min) & $\pm 1$ & $\pm 1$ & $\pm 2$ \\ 
$E_{R}$ ($10^{26}$ ergs) & $2.0^{+1.5}_{-0.7}$ & $3.1^{+0.8}_{-0.5}$ & $7.7^{+2.3}_{-2.1}$ \\ \hline
\end{tabular}
\end{center}
\renewcommand{\baselinestretch}{1.5}
\caption[]{A summary of the model parameters characterizing each of the observed flares.  The 
flares were modeled using the equation:
$F(t)=F_0e^{(t-t_0)/\tau_1}+F_q$ for $t<t_0$ and $F(t)=F_0e^{-(t-t_0)/\tau_2}+F_q$ for $t\ge t_0 \label{eqn:fit}$,
where $F_0$ is the peak flux density, $F_q$ is the quiescent flux of the 
source, and $\tau_1$ and $\tau_2$ are the rise and decay times, respectively.  
With this approach $F_0$, $t_0$, $\tau_1$, $\tau_2$, and $F_q$ are free parameters, and the total value of 
$\chi^2$ for all three flares is $3.7$ for 9 degrees of freedom.  Forcing 
$F_q$ to have the same value at all times, we find that the level of quiescent emission derived from the model-fit 
is consistent with the quiescent flux densities measured during observations of non-flaring emission; from  
modeling we find an average quiescent flux of $100\pm 35$ $\mu$Jy, 
while direct observations of the quiescent emission give a value of $75 \pm 23$ $\mu$Jy. 
\newline
The energy released in the flares at radio wavelengths is calculated using the equation
$E_R=4\pi d^2\nu_{\rm peak} F_{\nu {\rm peak}} \tau$, where $d=5$ pc is the distance to the source, 
$\nu_{\rm peak}\approx 8.5\times 10^9$ Hz is the peak frequency, $F_{\nu {\rm peak}}$ is the peak flux density
in units of erg sec$^{-1}$ cm$^{-2}$ Hz$^{-1}$, and $\tau=\tau_1+\tau_2$ is the temporal extent of 
the flare.  We find that all three flares released approximately the same amount of energy at 8.5 GHz.}
\label{tab:lc}
\end{table}

\clearpage
\begin{figure}[t!]
\centerline{\psfig{figure=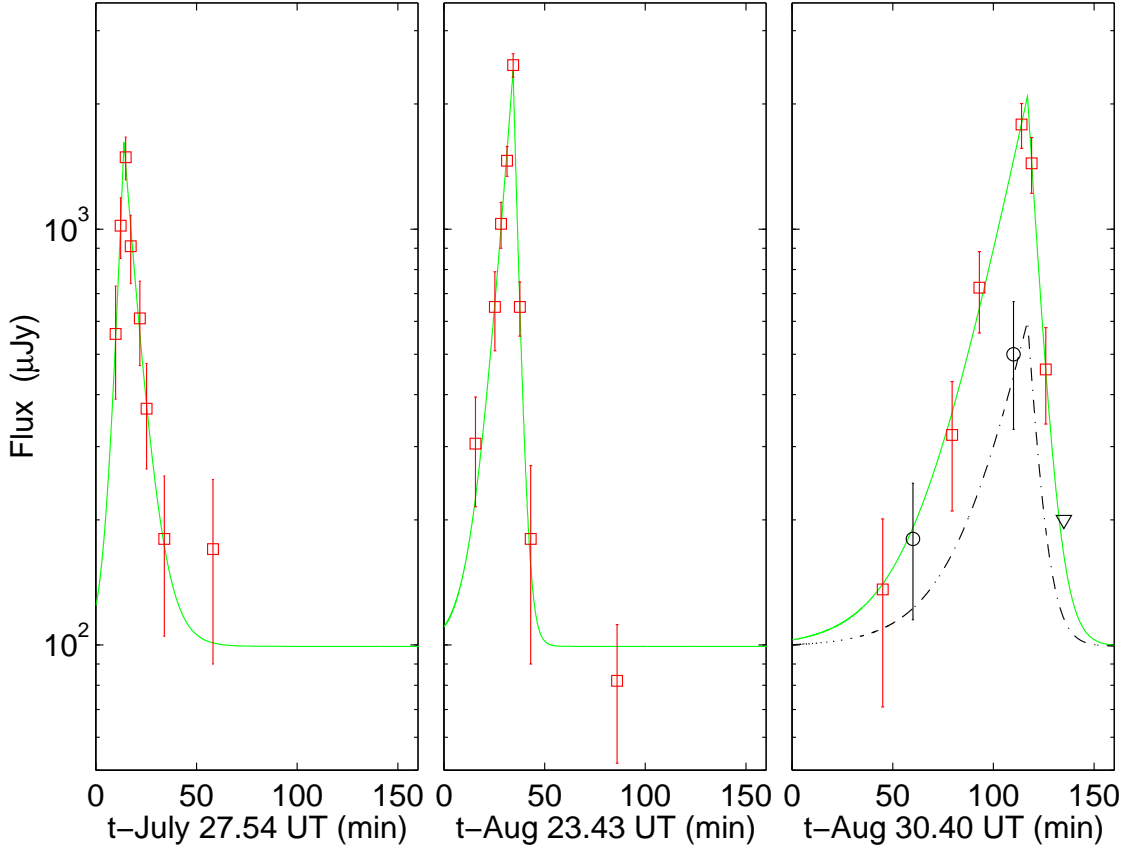,width=15cm}}
\caption[]{Light curves for the flaring radio emission from
LP944$-$20 at 8.5 GHz (squares) showing strong flares, quiescent emission,
and the self-absorbed emission at 4.9 GHz (circles; upper limit is given
by an inverted triangle).  The three panels are the lightcurves on (a)
2000 July 27.54, (b) 2000 August 23.43, and (c) 2000 August 30.40.  Time in minutes is relative
to the starting time of each observation, and the solid line is the model fit based on the fitting equation
given in Table~\ref{tab:lc}. Panel (c) also shows the flux density at 4.9 GHz,
which was measured simultaneously with the 8.5 GHz observation.  Based on the ratio of the flux densities at
these
two frequencies we find that the spectral index is $2.1\pm 0.3$, which indicates that the emission at 4.9 GHz 
is
self-absorbed.  The dashed line is a model fit based on the simple model (see notes Table~\ref{tab:lc}) and
the
spectral shape of optically-thick synchrotron emission, $F_\nu \propto \nu^{5/2}$.
\newline
Based on these lightcurves we can calculate the magnetic field strength and the number
density of trapped electrons using the equations for synchrotron emission:
$\frac{eB}{m_ec}\gamma_e^2 = \nu\approx 8.5 {\rm GHz}$ and
$\frac{4}{3}\sigma_Tc\gamma_e^2\frac{B^2}{8\pi}N = 4\pi d^2 \nu F_{\nu {\rm peak}} \approx 5\times 10^{23}$
erg
sec$^{-1}$, where $e$ and $m_e$ are the electron charge and mass, respectively, $\sigma_T$ is the Thomson
cross-section, $N$ is the total number of trapped electrons, $d$ is the distance to the source,
and $F_{\nu {\rm peak}}$ is the peak-flare flux density.
}
\label{fig:lc}
\end{figure}

\end{document}